\documentclass[prl,superscriptaddress,twocolumn,showpacs,preprintnumbers,amsmath,floatfix]{revtex4}
\usepackage{amssymb}

\usepackage{graphicx}

\begin{document}

\title{Can the Mott Insulator TiOCl be Metallized by Doping?\\A First-Principles Study.}

\author{Yu-Zhong Zhang}
\affiliation{Institut f\"ur Theoretische Physik, Goethe-Universit\"at
Frankfurt, Max-von-Laue-Stra{\ss}e 1, 60438 Frankfurt am Main, Germany}
\author{Kateryna Foyevtsova}
\affiliation{Institut f\"ur Theoretische Physik, Goethe-Universit\"at
Frankfurt, Max-von-Laue-Stra{\ss}e 1, 60438 Frankfurt am Main, Germany}
\author{Harald O. Jeschke}
\affiliation{Institut f\"ur Theoretische Physik, Goethe-Universit\"at
Frankfurt, Max-von-Laue-Stra{\ss}e 1, 60438 Frankfurt am Main, Germany}
\author{Martin U. Schmidt}
\affiliation{Institut f\"ur Anorganische und Analytische Chemie,
Goethe-Universit\"at Frankfurt, Max-von-Laue-Stra{\ss}e 7, 60438
Frankfurt am Main, Germany}
\author{Roser Valent{\'\i}}
\affiliation{Institut f\"ur Theoretische Physik, Goethe-Universit\"at
Frankfurt, Max-von-Laue-Stra{\ss}e 1, 60438 Frankfurt am Main, Germany}

\date{\today}

\begin{abstract}
We investigate the effect of Na intercalation in the layered Mott
insulator TiOCl within the framework of density functional theory.
We show that the system remains always insulating for all studied Na
concentrations, and the evolution of the spectral weight upon Na
doping is consistent with recent photoemission experiments. We
predict the Na-doped superlattice structures, and show that
substitutions of O by F, Cl by S, or Ti by  V (or Sc), respectively,
fail to metallize the system. We propose a description in terms of a
multiorbital ionic Hubbard model in a quasi-two-dimensional lattice
and discuss the nature of the insulating state under doping.
Finally, a likely route for metallizing TiOCl by doping is proposed.
\end{abstract}


\maketitle

A successful route to drive a layered compound into a (usually
unconventional) superconducting state is by doping the system with
additional electrons or
holes~\cite{Eisaki,Tokura,Kamihara,Rotter,Mizuguchi,Yamanaka,Takada}.
In contrast to the cuprates~\cite{Eisaki,Tokura}, Fe-based systems
~\cite{Kamihara,Rotter,Mizuguchi}, and cobaltates~\cite{Takada}
where 3d ions with occupations larger than 5 electrons are
responsible for superconductivity, in the layered nitride
superconductors~\cite{Yamanaka}, Zr and Hf are in 4d$^2$ and 5d$^2$
configurations. In this work we  investigate whether unconventional
superconductivity can be realized in early 3d layered compounds upon
doping. A good candidate is TiOCl which is, similar to the cuprates, a layered Mott insulator
at room temperature~\cite{Seidel} but with Ti$^{3+}$
in its 3d$^1$ configuration; Ti is in the same column of the
periodic table as Zr and Hf. According to these features, a
fundamental question is whether TiOCl can be a metal or even
superconductor by doping.

In fact, the undoped TiOCl, formed by stacked bilayers of TiO along
the $c$ axis separated by Cl ions, has already attracted
considerable
interest~\cite{Shaz,Rueckamp,Hoinkis,Abel,ZhangJeschkeValenti1} due
to the appearance of a zero-field structural incommensurate state.
Furthermore, great effort has been put into metallizing the undoped
TiOCl by applying hydrostatic
pressure~\cite{Kuntscher,ZhangJeschkeValenti2}.

Very recently, samples of TiOCl with Na intercalated into the
interstices of the TiO bilayers were prepared in order to drive the
system into a metallic or even superconducting state upon electron
doping~\cite{Sing}. However, photoemission experiments
(PES) show that a pronounced gap remains at the Fermi
level for all concentrations of Na, indicating that the insulating
state always survives electron doping~\cite{Sing}.  In view of these results, it
is essential to understand the microscopic origin of this puzzling
insulating state upon doping and, thereafter, find a more promising
recipe for metallizing TiOCl. Moreover, considerable interest in
insulator-insulator transitions~\cite{Fabrizio}
 motivates us to investigate the nature
of the doped insulating state.

\begin{figure}[tb]
\includegraphics[width=0.45\textwidth]{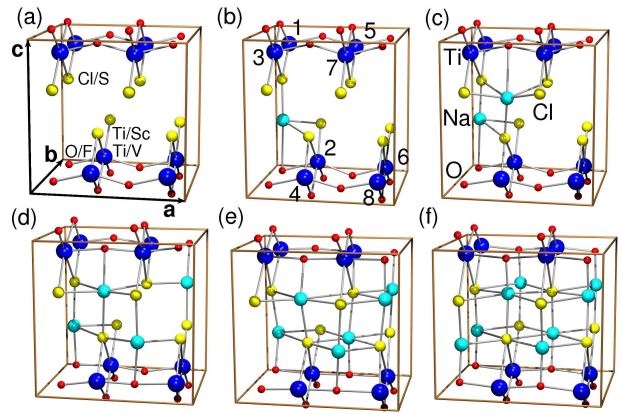}
\caption{(Color online) Superlattices at 6 different concentrations
of
  Na doping. (a) undoped case.   The atoms
  where substitutions of Cl by S, O by F and Ti by V(Sc) are performed  are
also specified. (b) 12.5 $\%$ Na doping:
  Na$_{1/8}$TiOCl. The 8  Ti ions in the  cell are labelled by numbers
  1 to 8. (c) 25 $\%$ doping: Na$_{1/4}$TiOCl, (d) 50 $\%$
  doping: Na$_{1/2}$TiOCl, (e) 75 $\%$  doping: Na$_{3/4}$TiOCl, and
  (f) 100 $\%$ doping: NaTiOCl. }
\label{fig:Superlattice}
\end{figure}

In this Letter, we performed {\it ab initio} molecular dynamics
studies with the Car-Parrinello~\cite{CarParrinello}
projector-augmented wave (CP-PAW) method~\cite{Bloechl} to study the
effect of Na doping on TiOCl within the Parrinello-Rahman
scheme~\cite{Rahman}. The Perdew-Burke-Ernzerhof generalized
gradient approximation (GGA)+$U$ to the density functional theory
(DFT) is employed in order to have the correct insulating behavior
and lattice structure in the undoped
case~\cite{ZhangJeschkeValenti1,ZhangJeschkeValenti2,Pisani,comment,comment2}.

Our findings can be summarized as follows: For all Na concentrations
studied, the unusual insulating state persists and the change of
spectral weight at elevated Na concentration agrees well with the PES
results~\cite{Sing}. Analysis of this insulating state shows that the
apropriate microscopic model is no longer a dominantly
quasi-one-dimensional single-band model valid in the undoped
case~\cite{Seidel,Hoinkis,Abel,ZhangJeschkeValenti1} but a
two-dimensional multi-orbital model. The effect of Na doping is
twofold: (i) it strongly modifies the crystal field splitting of Ti d
states by the Na$^+$ ion and (ii) it induces a coexistence of
Ti$^{3+}$ and Ti$^{2+}$ ions which is further stabilized by
long-range Coulomb interaction as usually observed in mixed-valence
compounds~\cite{ZhangFuldeThalmeierYaresko}.  Together with the
on-site Coulomb repulsion as well as the Hund's rule coupling, these
effects play a crucial role in forming the doped symmetry-breaking
insulating phase.  We further predict the Na-doped superlattice
structures, and we also find that doping TiOCl by substituting O by F,
Cl by S, or Ti by V (or Sc) does not result in a metal. While the
insulating state of the F/S substitution case can be ascribed to the
same microscopic origin as in the Na-doped case, the long-range
Coulomb interactions play a minor role in the V/Sc substitution.

For our calculations we used supercells obtained by doubling the
primitive unit cell (space group $Pmmn$, two formula units per unit
cell) in both $a$ and $b$ directions. The molecular formula under
doping is defined as Na$%
_x$Ti$_{1-y}$V$_y$(Sc$_y$)O$_{1-z}$F$_z$Cl$_{1-w}$S$_w$, where $x$,
$y$, $z$, and $w$ are the doping concentrations
(Fig.~\ref{fig:Superlattice} (a)).  We performed carefully converged
relaxations of lattice parameters and atomic positions
for each doping case with high energy cutoffs of 45 Ry and 180 Ry for
the wave functions and charge density expansion, respectively.

\begin{figure}[tb]
\includegraphics[angle=-90,width=0.45\textwidth]{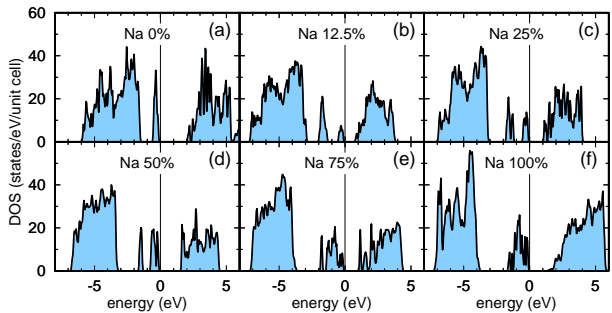}
\caption{(Color online) Total density of states (DOS) at the 6
different concentrations of Na doping shown in
Fig.~\protect\ref{fig:Superlattice}. (a) undoped case:
  TiOCl, (b)  Na$_{1/8}$TiOCl, (c)  Na$_{1/4}$TiOCl, (d)  Na$_{1/2}$TiOCl, (e)
   Na$_{3/4}$TiOCl, and (f)
  NaTiOCl. }
\label{fig:DOSNaAll}
\end{figure}

We first study the case of Na$_x$TiOCl for six different values
$x$=0, 1/8, 1/4, 1/2, 3/4, 1. Since no experimental data for the Na
positions are available, numerous possibilities have to be tested by
\textit{ab initio} molecular dynamics. However, according to
Pauling's rule that the coordination number is determined by the
radius ratio~\cite{Pauling}, the most probable Na positions are in
the cages of 5 Cl and 1 O as shown in Figs.~\ref{fig:Superlattice}
(b)-(f).  In fact, by performing \textit{ab initio} molecular
dynamics starting from several different initial Na positions, we
find that the Na atoms always fall into this cage to reach the
coordination number of 6. The remaining uncertainty is given by the
various combinations of locating Na atoms into the 8 cages of the
supercell.  We performed \textit{ab initio} molecular dynamics for
all the different combinations at various doping concentrations and
compared the total energies. Fig.~\ref{fig:Superlattice} presents
the final stable lattice structures with lowest total energies for
the six Na concentration values $x$ studied~\cite{disorder}. The
 supercell structures as a function of Na doping are
 shown in the figure and call for further experimental X-ray diffraction
studies.

In Fig.~\ref{fig:DOSNaAll} we present the total density of states
(DOS) for the six Na-doped TiOCl cases considered.  The DOS of the
undoped case (Fig.~\ref{fig:DOSNaAll} (a)) is correctly reproduced
with an energy gap around 2 eV, and the peak close to the Fermi
level is dominantly of $d_{x^2-y^2}$ character. Here we chose the
local coordinate frame as $x\parallel b$, $y\parallel c$ and
$z\parallel a$ with $d_{x^2-y^2}$, $d_{xz}$, $d_{yz}$ forming the
$t_{2g}$ bands and $d_{xy}$, $d_{z^2}$ the $e_g$ bands. With Na
doping (Figs.~\ref{fig:DOSNaAll} (b)-(f)) the whole spectral weight
is suddenly shifted towards lower energies, and an additional peak
with weight $2x$
appears close to the Fermi level
as compared to the undoped case.  The separation of the two Ti peaks
is about 1~eV for $x = 1/8$ and a gap always persists at the Fermi
level in all Na-doped cases. All the findings are consistent with
the experimental observations~\cite{Sing}.

\begin{figure}[tb]
\includegraphics[width=0.45\textwidth]{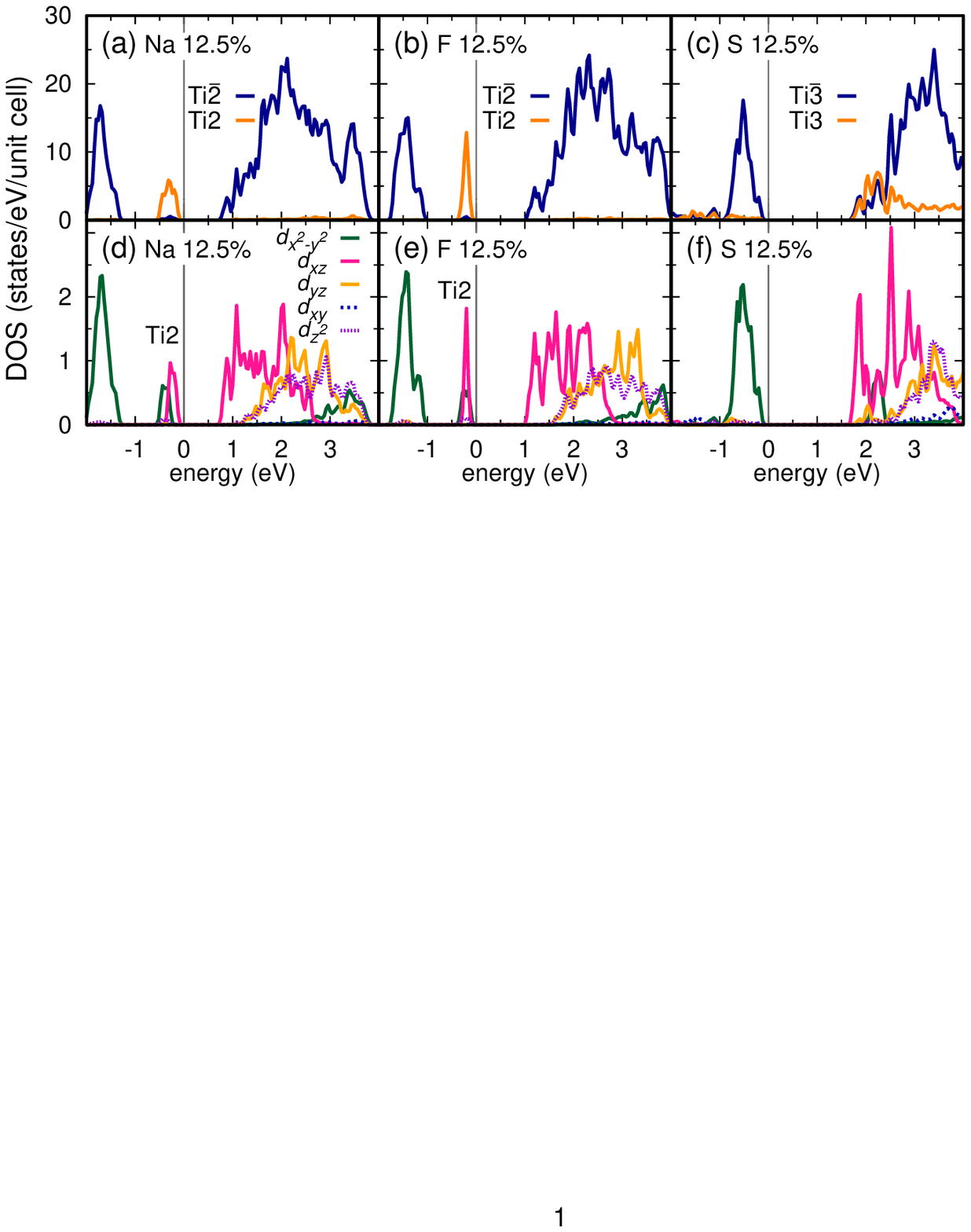}
\caption{(Color online) Partial DOS for (a), (d) 12.5$\%$ Na doping:
  Na$_{1/8}$TiOCl, (b), (e) 12.5$\%$ F substitution:
  TiO$_{7/8}$F$_{1/8}$Cl, and (c), (f) 12.5$\%$ S substitution:
  TiOCl$_{7/8}$S$_{1/8}$. In (a), (b), and (c) the total Ti-resolved
  DOS are shown. The grey (orange) curve denotes the DOS for the Ti
  ion nearest to the doping ion (Ti2 for the Na- and F-doped cases,
  Ti3 for the S-doped case).  The black (blue) curve denotes the DOS
  for the rest of Ti ions. In (d), (e), and (f) the corresponding
  orbital-resolved DOS are shown. Below $E_f$ the contributions of Ti2
  for the electron-doped cases are explicitly marked.}
\label{fig:pDOSNaFS}
\end{figure}

In order to understand the microscopic origin of this doped
insulating state, we show in Figs.~\ref{fig:pDOSNaFS}~(a) and (d)
the orbital-resolved DOS for a Na doping of $x$=1/8.  Close to the
Fermi level, both $d_{x^2-y^2}$ and $d_{xz}$ orbitals contribute to
the DOS. This can be understood by performing a rough energy
estimation in the atomic limit when one additional electron is added
to the Ti t$_{2g}$ bands. While the change of total energy is
$U+2J_\text{H}$ if the additional electron occupies the
$d_{x^2-y^2}$ orbital, it is $U-J_\text{H}+\Delta$ if the electron
occupies the $d_{xz}$ orbital where $\Delta$ is the orbital
excitation from $d_{x^2-y^2}$ to $d_{xz}$, and $J_\text{H}$ is the
Hund's rule coupling. Since $\Delta$ $\sim$ 0.1~eV and
$J_\text{H}\sim 1$~eV, adding the additional electron to the
$d_{xz}$ orbital is always preferable. This analysis indicates the
importance of the Hund's rule coupling in the doping case.

The presented DFT results point to a multi-band scenario.  With the
help of $N$-th order muffin tin orbital (NMTO)
downfolding~\cite{tanusri} we found that the hopping integrals
within $d_{xz}$ orbitals along $a$ and within one TiO bilayer  are
comparable to the hopping integrals within $d_{x^2-y^2}$ orbitals
along $b$, indicating that the interaction network is
two-dimensional under doping.  Since the exchange interactions among
local spins along $a$ and within one TiO bilayer are weakly
ferromagnetic~\cite{ZhangJeschkeValenti1}, the doped electron would
in principle hop freely  in the  system ~\cite{Zener} and would
result in a metallic state.  This is though in contradiction with
the insulating state obtained both in our DFT calculations and in
experimental observations.

\begin{figure}[tb]
\includegraphics[angle=-90,width=0.45\textwidth]{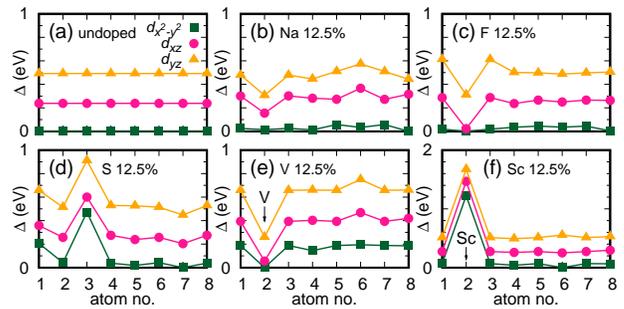}
\caption{(Color online) t$_{2g}$ crystal field splitting energies
obtained from GGA calculations.
   The horizontal axis denotes 8 Ti ions in the supercell
  labelled by numbers from 1 to 8 as shown in
  Fig.~\ref{fig:Superlattice} (b). (a) TiOCl, (b) Na$_{1/8}$TiOCl, (c) TiO$_{7/8}$F$_{1/8}$Cl, (d)
TiOCl$_{7/8}$S$_{1/8}$, (e) Ti$_{7/8}$V$_{1/8}$OCl, and (f)
Ti$_{7/8}$Sc$_{1/8}$OCl. } \label{fig:CFS}
\end{figure}

Therefore, to further elucidate the possible mechanism for the
insulating state upon doping, we compare the crystal field splitting
of the undoped case with that of the 1/8-Na-doped case as shown in
Figs.~\ref{fig:CFS}~(a) and (b). In the undoped case all 8 Ti ions
in the supercell are equivalent and the orbital excitations are the
same. Here, the Ti ions are labelled from Ti1 to Ti8 as shown in
Fig.~\ref{fig:Superlattice}~(b). In the 1/8-Na-doped case, all Ti
ions become inequivalent and the crystal field splittings are
different from site to site. Most importantly, the splittings of
t$_{2g}$ orbitals on the Ti2 (closest Ti to Na) become strikingly
small, which makes it possible to trap the additional electron into
$d_{xz}$ orbitals on this Ti site. This finding is consistent with
the partial DOS shown in Figs.~\ref{fig:pDOSNaFS}~(a) and (d) where
the peaks near the Fermi level are mainly contributions from the
$d_{x^2-y^2}$ and $d_{xz}$ orbitals of Ti2.  The peak at lower
energies is from the rest of Ti ions (all denoted by Ti${\bar{2}}$)
and is of purely $d_{x^2-y^2}$ character. Further investigations of
the crystal field splittings and partial DOS for the other Na doping
concentrations reveal that the same mechanism can be applied to the
observed insulating states, {\it i.e.}, each Na ion strongly
modifies the lattice structure locally, reduces the crystal field
splitting of the t$_{2g}$ orbitals on the closest Ti ion and leads
to localization of the doped electron, which reduces the Ti$^{3+}$
($3d^1$) to Ti$^{2+}$ ($3d^2$) and prevents conduction. Furthermore,
long-range Coulomb repulsion becomes effective due to the appearance
of mixed-valence state of Ti$^{3+}$ and Ti$^{2+}$ and further
stabilizes the symmetry-breaking insulator. Within the point-charge
approximation~\cite{Sing}, the combination of crystal field
splitting and the long-range Coulomb interaction can account for the
gap of 1eV.

We have also considered the effect of substituting O by F
(Fig.~\ref{fig:pDOSNaFS} (b), (e)), Cl by S (Fig.~\ref{fig:pDOSNaFS}
(c), (f)), and Ti by V or Sc (not shown) in the undoped TiOCl at
concentrations of $1/8$, leading to TiO$_{7/8}$F$_{1/8}$Cl,
TiOCl$_{7/8}$S$_{1/8}$, Ti$_{7/8}$V$_{1/8}$OCl and
Ti$_{7/8}$Sc$_{1/8}$OCl.  The atomic positions where the
substitutions are made (without loss of generality) are specified in
Fig.~\ref{fig:Superlattice} (a). In all the doped cases gaps still
open at the Fermi level. The DOS of the F-doped case
(Fig.~\ref{fig:pDOSNaFS} (b), (e)) is similar to that of the
Na-doped case with a double peak structure below the Fermi level,
since both are electron doping processes. The similarities can be
also observed in the crystal field splittings
(Fig.~\ref{fig:CFS}~(c)). However, in the F-doped case the Ti2
$d_{x^2-y^2}$ and $d_{xz}$ orbitals are almost degenerate.

Substituting Cl by S implies taking out an electron from the
$d_{x^2-y^2}$ orbital (hole doping), and one would expect the system
to be metallic.  However, from Fig.~\ref{fig:CFS}~(d), we find that
the on-site orbital energies of Ti3 (Fig.~\ref{fig:CFS}~(d)) which
is closest to the substituted S ion are significantly raised due to
the large distortion of the lattice structure, again leading to a
localized state for the doped hole.  In this case, no spectral
weight from the Ti3 $3d$ orbitals is detected close to the Fermi
level (Fig.~\ref{fig:pDOSNaFS} (c)).  Interestingly, the Ti1
$d_{x^2-y^2}$ orbital energy is almost degenerate with the Ti7
$d_{xz}$ orbital energy, indicating the importance of on-site
Coulomb interaction which avoids double occupation on each Ti ion.
This leads to a DOS  below the Fermi level of only $d_{x^2-y^2}$
character (Fig.~\ref{fig:pDOSNaFS} (f)).  On the other hand, doping
with S might result in the formation of S$_2^{2-}$ ions, which are
isoelectronic to two Cl$^-$ ions and do not contribute to doping.
The system will then remain insulating.  Finally, substituting Ti by
V (electron doping) and Sc (hole doping) we observe a similar effect
as seen in the previous electron- and hole-doped cases. The onsite
orbital energies are much lower (higher) on the V (Sc) ion than on
the Ti ion (Figs.~\ref{fig:CFS}~(e) and (f)), which can again
account for the localization of the doped electron (hole) and the
absence of a metallic state. However, the oxidations are all $+3$
(Ti$^{3+}$, V$^{3+}$, Sc$^{3+}$),  indicating that the role of
long-range Coulomb interaction is negligible.

From the above analysis, we conclude that the appropriate
microscopic model that accounts for the persistence of the
insulating state upon doping is given by a multiband ionic
Hubbard model:
\begin{align}
H& =\sum_{i,j,\sigma ,\tau \tau^{\prime}}t_{ij,\tau
\tau^{\prime}}c_{i\tau \sigma }^{\dagger }c_{j\tau^{\prime} \sigma
}+\left( U+2J_\text{H}\right) \sum_{i,\tau }n_{i\tau \uparrow
}n_{i\tau
\downarrow }  \nonumber \\
& +U\sum_{i,\tau >\tau ^{\prime },\sigma }n_{i\tau \sigma }n_{i\tau ^{\prime
}\overline{\sigma }}+\left( U-J_\text{H}\right) \sum_{i,\tau >\tau ^{\prime
},\sigma }n_{i\tau \sigma }n_{i\tau^{\prime} \sigma }  \nonumber \\
& +\sum_{i,\sigma ,\tau }\Delta _{i,\tau }n_{i\tau \sigma
}+\sum_{i,j}V_{ij}n_in_j. \label{Eq:MIH}
\end{align}
The first term describes the hopping between Ti sites $i,j$ within
the three t$_{2g}$ orbitals ($\tau$, $\tau^{\prime}$), the second to
fourth terms are the intra- and inter-orbital Coulomb interactions
and the Hund's rule coupling, respectively, the fifth term the
crucial on-site orbital energies $\Delta_{i,\tau}$, and the last
term the effective long-range intersite Coulomb interaction
$V_{ij}$. Upon Na doping, $\Delta_{i,\tau}$ differs from site to
site which forms a local trapping potential and leads to a
coexistence of Ti$^{3+}$ and Ti$^{2+}$ stabilized by $V_{ij}$. Then,
a doping-induced phase transition from undoped Mott insulator to
doped insulator with charge disproportionation occurs. While a
similar transition has been extensively investigated in the one-band
case~\cite{Fabrizio}, inclusion of orbital degrees of freedom may
result in an even richer phase diagram~\cite{Medici,Lee}.

Although our DFT calculations are only performed for several
commensurate Na doping cases, we argue that the scenario for this
insulating state under doping can be applied to all doping
concentrations: the  Na ion enters an individual cage consisting of
5 Cl and 1 O ions and creates a trapping potential reducing the
crystal field splitting of the t$_{2g}$ orbitals on the closest Ti
ion and resulting in a localized state for the doped electron with
the help of long-range Coulomb interaction.

Finally, with the understanding of the mechanism for the insulating
state, we propose two ways to metallize TiOCl by doping which would
favor  a possible superconducting state in a doped Mott insulator
by suppressing other symmetry-breaking states. The first idea is to
avoid the strong modification of crystal field splitting and the
formation of a trapping potential due to the lattice deformation
induced by the cation or anion. Therefore, a possible direction is
to intercalate Na together with organic ligands to prevent the doped
Na from entering the cage, in the spirit of intercalating
organosolvated Li into $\beta$-HfNCl~\cite{Yamanaka}. A second
practicable way is to apply external
pressure~\cite{Kuntscher,ZhangJeschkeValenti2} after carrying out
electron doping. Due to the appearance of the additional peak and
the reduction of the gap under doping, the doped system becomes much
easier to metallize under pressure than the undoped one.

In conclusion, we performed DFT calculations on doped TiOCl and
unveiled the microscopic origin of the observed insulating state
upon doping.  The superlattice structures at different Na doping
concentrations and the effects of different substitutions are
predicted. We find that the opportunity to metallize TiOCl or even
to make it superconducting still remains. Moverover, the
insulator-to-insulator transition itself has been extensively
discussed by tuning the correlation strength~\cite{Fabrizio}.  Here,
we open a new way to induce the insulator-insulator transition by
doping. Finally, our proposed model may contain novel phases which
calls for further investigations by many-body
techniques~\cite{Medici,Lee}.

We thank R. Claessen, M. Sing, M. Knupfer, M. Abd-Elmeguid, and C.
Gros for useful discussions.  We thank the Deutsche
Forschungsgemeinschaft for financial support through the TRR/SFB~49,
FOR~412 and Emmy Noether programs, and we acknowledge support by the
Frankfurt Center for Scientific Computing.


\begin{thebibliography}{99}
\bibitem{comment} $U=4.5$~eV and $J_\text{H} = 1.0$~eV were chosen
  within GGA+$U$ to reproduce the correct gap of
  $2~eV$ for undoped TiOCl.
\bibitem{disorder} Other metastable configurations may coexist
with the groundstate one at finite temperature, which may account
for the soft gap observed experimentally at higher Na-doping. Here,
we only aim at the ground state.
\end{thebibliography}
\end{document}